\documentclass[acmsmall,screen]{acmart}

\setcopyright{none}
\renewcommand\footnotetextcopyrightpermission[1]{}
\usepackage{graphicx} 
\usepackage{float}
\usepackage{listings, listings-rust}
\usepackage{xcolor}
\usepackage{subfigure}
\usepackage{mdframed}
\usepackage[most]{tcolorbox}
\usepackage[most]{tcolorbox}
\usepackage{algorithm}
\usepackage{algpseudocode}
\usepackage{multirow}
\usepackage{hyperref}
\definecolor{lightred}{rgb}{1.0, 0.8, 0.8}
\definecolor{lightgreen}{rgb}{0.8, 1.0, 0.8}
\definecolor{gray}{rgb}{0.9, 0.9, 0.9}

\usepackage[dvipsnames]{xcolor}
\definecolor{codegreen}{rgb}{0,0.6,0}
\definecolor{codegray}{rgb}{0.5,0.5,0.5}
\definecolor{codepurple}{rgb}{0.58,0,0.82}

\newcommand{\CRust}{\textsc{C2Rust}}
\newcommand{\CSaferRust}{\textsc{C2SaferRust}}
\newcommand{\TranslationGym}{\textsc{TranslationGym}}

\newcommand{\HumanWritten}{Human-Written}

\author{Biruk Tadesse}
\affiliation{%
  \institution{North Carolina State University}
  \country{USA}
}
\email{batadess@ncsu.edu}

\author{Vikram Nitin}
\affiliation{%
  \institution{Columbia University}
  \country{USA}
}
\email{vikram.nitin@columbia.edu}

\author{Mazin Salah}
\affiliation{%
  \institution{North Carolina State University}
  \country{USA}
}
\email{mbsalah@ncsu.edu}

\author{Baishakhi Ray}
\affiliation{%
  \institution{Columbia University}
  \country{USA}
}
\email{rayb@cs.columbia.edu}

\author{Marcelo d'Amorim}
\affiliation{%
  \institution{North Carolina State University}
  \country{USA}
}
\email{mdamori@ncsu.edu}

\author{Wesley K. G. Assunção}
\affiliation{%
  \institution{North Carolina State University}
  \country{USA}
}
\email{wguezas@ncsu.edu}

\begin{document}

\title{Code Quality Analysis of Translations from C to Rust}

\begin{abstract}
C/C++ is a prevalent programming language. Yet, it suffers from significant memory and thread-safety issues. Recent studies have explored automated translation of C/C++ to safer languages, such as Rust. However, these studies focused mostly on the correctness and safety of the translated code, which are indeed critical, but they left other important quality concerns (e.g., performance, robustness, and maintainability) largely unexplored.  Such an understanding is essential not only for improving translation techniques but also for helping developers assess trade-offs among safety, scalability, and long-term viability.
This work investigates strengths and weaknesses of three C-to-Rust translators, namely \CRust\ (a transpiler), \CSaferRust\ (an LLM-guided transpiler), and \TranslationGym\ (an LLM-based direct translation).
We perform an in-depth quantitative and qualitative analysis of several important quality attributes for the translated Rust code of the popular GNU coreutils, using human-based translation as a baseline.
To assess the internal and external quality of the Rust code, we: (i) apply Clippy, a rule-based state-of-the-practice Rust static analysis tool; (ii)~investigate the capability of an LLM (GPT-4o) to identify issues potentially overlooked by Clippy; and (iii) perform a manual analysis of the issues reported by Clippy and GPT-4o.
Our results show that while newer translation techniques (e.g., \TranslationGym) reduce some unsafe and non-idiomatic patterns, they frequently introduce new issues (e.g., runtime panic risks, performance regressions, and thread-safety concerns), revealing systematic trade-offs that are not visible under existing evaluation practices. Notably, none of the automated techniques consistently match or exceed human-written translations across all quality dimensions, yet even human-written Rust code exhibits persistent internal quality issues such as readability and non-idiomatic patterns. Together, these findings show that translation quality remains a multi-dimensional challenge, requiring systematic evaluation and targeted tool support beyond both naive automation and manual rewriting.
As contributions, this paper provides a taxonomy of translation issues, an in-depth evaluation of three existing translators, and actionable guidance to help tool builders and developers design and assess more trustworthy translations.
\end{abstract}




\maketitle

\section{Introduction}
\label{sec:introduction}

C and C++ are widely used programming languages for critical software~\cite{TIOBE2026}, such as operating systems and database management systems~\cite{Popescu2025, Pereira2022}. However, it is widely known that C/C++ suffer from significant memory issues because of their lack of memory and thread safety~\cite{shetty2019crust,durumeric2014}, which have led to critical bugs and security flaws~\cite{durumeric2014, leveson1993}. 
Given these memory issues, the industry~\cite{hong2025}, government~\cite{WhiteHouse2021}, and open-source projects~\cite{uutils2025} have been translating legacy systems from C/C++ to Rust, which is explicitly designed to eliminate these risks~\cite{klabnik2023rust,Emre2021}. 

Manually translating large codebases from C/C++ to Rust is labor-intensive and error-prone~\cite{Li2025}. Microsoft has been rewriting the Windows core from C++ to Rust since 2020, with no estimated completion date~\cite{Thomas2023}.
As a consequence, recent studies have explored automated techniques to translate C/C++ to Rust~\cite{C2Rust2018, C2SaferRust2025, TranslationGym2025, Mehmet2023, Hanliang2023}.
The main focus of these techniques is addressing memory issues, with the translated Rust code evaluated only for correctness and safety~\cite{C2Rust2018, C2SaferRust2025, Mehmet2023, Hanliang2023}. Unfortunately, these studies leave unexplored critical quality aspects of the translated code, such as performance, robustness, and maintainability. 
In software development, code quality involves trade-offs among different properties (e.g.,  scalability and long-term viability) that extend beyond only functional correctness and safety~\cite{spinellis2006code,Simoes2024,yetistiren2023,damasceno2023test,lee2022engineering}. 

The goal of our work is to systematically study the strengths and weaknesses of three translation techniques that adopt different strategies: \CRust\ (a transpiler), \CSaferRust\ (an LLM-guided transpiler), and \TranslationGym\ (an LLM-based direct translation). 
We perform an in-depth quantitative and qualitative analysis of the translated Rust code for seven programs from the GNU coreutils~\cite{coreutils2025} that were translated using the techniques under investigation. Additionally, the human translations for these seven programs serve as a baseline. 
Our methodology to assess the quality of the Rust code relies on: (i) applying Clippy~\cite{li2024clippy}, a rule-based state-of-the-practice Rust static analysis tool; (ii) leveraging an LLM (GPT-4o) as a linter to identify issues potentially overlooked by Clippy; and (iii) performing detailed manual analysis of the issues returned by Clippy and GPT-4o.
To compare techniques, we proposed a taxonomy with 18 categories of quality dimensions that enable robust evaluation of Rust code not for correctness and safety (i.e., external quality), but also for future maintenance and evolution of the translated programs (i.e., internal quality).


Our empirical results show that the translation techniques exhibit systematic and non-obvious trade-offs across internal and external quality dimensions. Improvements in one aspect of quality often come at the expense of another. While more advanced approaches that build on transpilers and LLMs can reduce unsafe memory usage or non-idiomatic constructs, they frequently introduce new issues as runtime panic risks, performance, thread safety, and error handling. Yet, despite the advancements with LLMs, no automated technique consistently dominates across all quality dimensions. Notably, human-written Rust translations perform better in critical safety-related dimensions but still exhibit substantial internal quality issues, particularly in readability, documentation, and idiomaticity. 
Our results also indicate that Clippy, despite its widespread use in practice, has limitations when assessing translated, non-idiomatic Rust code. While its rule-based approach is effective for idiomatic Rust, it struggles to detect certain critical issues in C-style translations created by the transpiler. 
These findings demonstrate that translation quality is inherently multi-dimensional and remains challenging even for expert developers, underscoring the need for systematic evaluation frameworks and targeted guidance to support both tool builders and practitioners.

The contributions of our work are threefold: (i) we propose an \textit{empirical taxonomy} of 18 categories of Rust quality issues arising from C translation, offering a structured framework for future research and tool evaluation; (ii) we present the \textit{first systematic comparison} of three automated translation techniques (i.e., \CRust, \CSaferRust, \TranslationGym) against a \HumanWritten\ baseline; and (iii) we provide \textit{insights and actionable recommendations} for tool builders and developers on how to design safer, more idiomatic, and maintainable translations, based on our qualitative analysis of recurring pitfalls (e.g., preserving unsafe boundaries and avoiding runtime panics).

\section{Study Design}
\label{sec:study}

\begin{figure}[!tp]
    \centering
    \includegraphics[width=.9\linewidth]{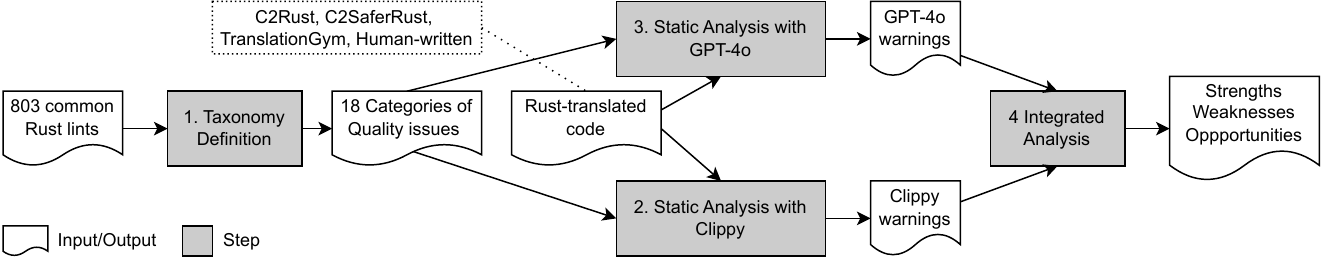}
    \caption{Study Design Steps}
    \label{fig:study_design}
\end{figure}

The goal of our study is to \textit{evaluate the internal and external quality of Rust code translated from C/C++}. 
To achieve this goal, our study is guided by the following research question (RQ):

\vspace{2mm} \noindent
\textbf{RQ. How does the quality of Rust code compare for different translation techniques, and what insights do these differences provide for improving the techniques?} 
This RQ aims to compare the quality of Rust code translated using three automated techniques and human-based translations. To enable this comparison, we define a taxonomy of 18 quality issue categories. By answering this RQ, we can identify the strengths and weaknesses of each technique, providing insights into what needs improvement for better translation. 
Figure~\ref{fig:study_design} presents the main phases for the systematic quantitative and qualitative analysis to answer our study RQ. The next sub-sections present more details for each phase.

\subsection{Taxonomy Definition: Rust Issue Categories}
\label{sec:categories}


To enable comparison of code quality across translation techniques, we create a taxonomy of Rust quality issue categories. Table~\ref{tab:categories} presents the 18 categories designed to capture a comprehensive range of internal and external quality properties relevant to translated code. 

The source of information to create this taxonomy was the list of 803 lint rules\footnote{\url{https://rust-lang.github.io/rust-clippy/master/index.html}} available on Clippy. This list provides a comprehensive, documented, and community-maintained enumeration of recurring Rust quality pitfalls, making it an empirically grounded and reproducible basis for deriving taxonomy categories. We began by manually reviewing all 803 lint rules to understand common Rust code issues. Clippy lints are accompanied by detailed documentation, including their purpose and illustrative examples, which makes their intent explicit and supports consistent interpretation.
We used Clippy's default groupings (e.g., style, complexity, correctness) as a reference point, but refined and expanded them to better capture the details relevant to the translated code. The categorization was performed through an iterative process, following principles of thematic analysis~\cite{cruzes2011}. 
We first drafted an initial set of categories (open code), then refined them as patterns emerged across lints to ensure that each lint mapped to a single, most representative category (axial code). The first two authors conducted the initial categorization, after which the other authors reviewed and updated the assignments; any conflicts were discussed and refined jointly to reach consensus. Each lint was grouped into one of our custom-defined categories by selecting the one that best represented the issue's core nature. 
Additionally, to ground the taxonomy in established quality models, we aligned each category with the ISO/IEC 25010~\cite{ISO25010} quality characteristics where its impact is primarily observed. This alignment links our Rust-specific categories to general software quality concerns while preserving their relevance to the translation context. The detailed mapping is provided in the supplementary material~\cite{supplementary}.


\begin{table}[!tp]
\caption{Taxonomy with categories used to classify Rust code issues}
\label{tab:categories}
\addtolength{\tabcolsep}{-2pt}
\footnotesize
\centering
\begin{tabular}{l|l|p{10cm}}
\toprule
& \textbf{Category} & \textbf{Description} \\
\midrule
\multirow{8}{*}{\rotatebox[origin=c]{90}{Internal Quality}} & Convention violation & Code that violates common Rust naming and design conventions. \\
& Documentation issues & Issues in comments or documentation that reduce comprehensibility or maintainability. \\
& Inflexible code & Code that uses overly specific types, limiting reusability and flexibility. \\
& Misleading code & Code that leads readers to believe it does something other than what it actually does. \\
& Non-idiomatic code & Code that does not follow Rust conventions, patterns, or best practices. \\
& Non-production code & Code meant for debugging or placeholder purposes that should not appear in production. \\
& Readability issues & Code that makes it harder for readers to understand the developer's intention. \\
& Redundant code & Unnecessarily duplicated code that does not contribute new behavior or logic. \\
\midrule
\multirow{10}{*}{\rotatebox[origin=c]{90}{External Quality}} & Arithmetic issues & Patterns that can lead to bugs or undefined behaviors due to arithmetic operations. \\
& Attribute issues & Improper or missing use of Rust attributes that affect code behavior or stability. \\
& Compatibility issues & Code that may not work across platforms, Rust versions, or environments. \\
& Error handling issues & Code that handles errors but hides root causes or limits debuggability. \\
& Logical issues & Code with valid syntax but likely reflects a misunderstanding in logic. \\
& Memory safety & Code that risks dangling pointers, buffer overflows, use-after-free, or data races. \\
& Performance & Code that compiles and runs correctly but leads to inefficient execution. \\
& Runtime Panic risks & Code that may trigger a panic during execution due to unchecked operations. \\
& Thread safety & Code that may cause undefined behavior or data races when used across multiple threads. \\
& Type safety & Code that discards type guarantees. \\
\bottomrule
\end{tabular}
\end{table}

\subsection{Translation Techniques}
\label{sec:techniques}
Our study compares the Rust-translated code quality of three automated translation techniques, together with human-based translation. Each of these is described in what follows. 

\textbf{\CRust}~\cite{C2Rust2018} is a transpiler designed to mechanically convert C code into functionally equivalent Rust. Its primary goal is to preserve the original behavior and ensure test suites continue to pass. The generated Rust code closely mirrors the input C. 

\textbf{\CSaferRust}~\cite{C2SaferRust2025} is built upon \CRust. It first uses the \CRust\ transpiler to generate an initial base of unsafe Rust code, then leverages an LLM to decompose this code into smaller, manageable units (e.g., individual functions). Then, it incrementally rewrites them into safer, more idiomatic Rust, aiming to reduce the reliance on unsafe constructs while preserving correctness.

\textbf{\TranslationGym}~\cite{TranslationGym2025} is a framework for direct LLM-based translation that does not rely on an initial transpilation step. It automates the complex process of translating a full project by breaking it down into individual functions. For each function, it provides the LLM with the necessary context (e.g., dependencies, global variables, sample execution values) and then integrates the LLM's translation back into the project. It uses compiler feedback and test results in an iterative loop to repair and validate the translated code.

\textbf{\HumanWritten}~\cite{uutils2025} code serves as our baseline. These are manual translations by experienced Rust developers contributing to the uutils coreutils project, an open-source effort to reimplement the GNU coreutils in Rust. The developers aim to preserve GNU-compatible behavior while writing and maintaining the code directly in Rust without relying on automated translation tools.

\subsection{Quality Assessment Tools}
\label{sec:quality_tools}

The goal of our work is to evaluate the quality of Rust-translated code. For that, our study relies on multiple analyses based on: (i) \textit{Clippy}~\cite{li2024clippy}, a rule-based state-of-the-practice linter that applies static analysis; (ii)  \textit{GPT-4o~\cite{GPT4o2024}}, an LLM that identifies issues based on the understanding of the program context, which are potentially overlooked by Clippy; and (iii) \textit{manual analysis} to assess issues reported by Clippy and GPT-4o, according to the quality dimensions defined in our taxonomy (see Table~\ref{tab:categories}). Thus, we do not rely on a single analysis tool; instead, \textit{Clippy} and \textit{GPT-4o} are intentionally used as complementary evaluators. Then, in our manual analysis, we interpret results through this combined view rather than treating either tool as a definitive oracle.


\textbf{Clippy.}
Clippy is a widely used static analysis tool in the Rust ecosystem~\cite{li2024clippy}, providing many lints that help catch common mistakes and promote idiomatic code patterns. However, simply collecting lint warnings was insufficient to draw meaningful conclusions across different translation methods. 
Thus, we created our taxonomy (see Section~\ref{sec:categories}) to aggregate lint results at a higher semantic level and compare them across techniques more effectively. We applied Clippy to four versions of each program (see Table~\ref{tab:programs}).
After collecting the raw Clippy output, we extracted the lint names and mapped them to our custom categories. This produced structured output that could be visualized and compared across the different translation techniques.


While Clippy is powerful, its lints are designed with idiomatic Rust, often missing issues in C-style Rust code (code that compiles but is not idiomatic or safe). For example, lints like \texttt{uninit\_vec} (uninitialized Vec usage) rely on detecting patterns involving Rust's standard library types. In C-style Rust code that uses raw pointers or unsafe blocks instead of \texttt{Vec} or other safe abstractions, these lints may not trigger because the code does not use the expected idiomatic constructs. As a result, Clippy might miss potential issues (e.g., unsafe memory access or uninitialized memory) in translated or C-style Rust code, even though such code could be error-prone or unsafe.

\textbf{LLM.} Knowing Clippy's limitations, we explored the use of GPT-4o~\cite{GPT4o2024} as a general-purpose linter. Unlike Clippy, which relies on hard-coded rules, the LLM can reason about non-idiomatic patterns and potentially unsafe practices in a broader context. We created a prompt (available in supplementary material~\cite{supplementary}) that asks GPT-4o to analyze chunks of Rust code, identify potential issues, and assign them to one of our predefined 18 categories (see Table~\ref{tab:categories}). This allowed us to maintain a consistent comparison with Clippy's categorization.

\textbf{Manual Analysis.} Finally, to understand the reasons behind the observed warning distributions, we manually conducted a systematic qualitative analysis performed by two authors. This process was designed to uncover patterns explaining why certain techniques performed better or worse in specific categories and to identify cases where resolving one issue introduced another. 
The core of this analysis was a function-level comparison across translation techniques.
For each subject program, we selected representative functions based on the presence of warnings, coverage of different quality categories in the taxonomy, diversity of functionality and, for each selected function, we examined the warnings reported by the static analysis tools alongside the corresponding Rust implementations and, when needed, the original C version. This inspection focused on differences along the quality dimensions defined in our taxonomy.

\subsection{Subject Programs}

We use a dataset of seven C programs collected from GNU coreutils~\cite{coreutils2025}. Table~\ref{tab:programs} presents the name and number of lines of code (LOC)\footnote{LOC refers to literal lines of code, including source code, comments, and attributes. We adopt this definition because translation techniques differ in the amount of boilerplate, wrapper code, and annotations they introduce.} for the Rust-translated code.
These programs implement the basic file, shell, and text manipulation utilities of the GNU operating system, including \texttt{pwd} and \texttt{cat}. We chose coreutils because it is a large, mature C project, with many constituent programs of varying sizes. Further, there is an ongoing community effort to manually rewrite coreutils in Rust~\cite{uutils2025}, which means that translating coreutils to Rust is an important real-world problem. 
For these seven programs, we have complete Rust translations available for all four techniques under study. 
This allows us to immediately analyze these translations without the overhead of running each translation ourselves. Finally, each dataset in coreutils includes end-to-end system-level test cases carefully constructed using fuzzing to ensure high coverage.  

\begin{table}[!tp]
    \caption{Subject programs (LOC)}
    \label{tab:programs}
    \footnotesize
    \centering
    \begin{tabular}{l|rrrrr}
        \toprule
        \textbf{Program} & \textbf{\CRust} & \textbf{\CSaferRust} & \textbf{\TranslationGym} & \textbf{\HumanWritten}\\
        \midrule
        Cat & 7460 & 7280 & 2575 & 890 \\
        Head & 8047 & 7757 & 3244 & 1626 \\
        Pwd & 5859 & 5733 & 1879 & 163 \\
        Split & 13848 & 12993 & 4755 & 3301 \\
        Tail & 14423 & 13484 & 6954 & 3308 \\
        Truncate& 7181 & 6906 & 2378 & 432 \\
        Uniq & 8299 & 8290 & 1605 & 745 \\
        \bottomrule
    \end{tabular}
\end{table}


\subsection{Integrated Quantitative and Qualitative Analysis}


After obtaining the categorized Clippy and GPT-4o results, we conducted a quantitative analysis to understand how each translation method performs across different dimensions of code quality. We analyze the results in two ways, per tool and per program. At the \textit{tool level}, we aggregated lint frequencies by category for each translation technique to understand their overall strengths and weaknesses. For example, we could identify whether a given approach tends to produce code with more \textit{Memory Safety} warnings or more \textit{Non-idiomatic code}. At the \textit{program level}, we performed a more granular analysis, allowing us to observe whether certain programs were more prone to particular types of quality issues depending on the translation method used. Additionally, we applied non-parametric statistical tests to formally evaluate differences across translation techniques. We used the Friedman test~\cite{friedman1937use} to assess overall differences in warning distributions, followed by the Nemenyi post hoc test~\cite{nemenyi1963distribution} to identify which pairs of techniques differed significantly.
\section{Results}
\label{sec:results}



\begin{figure*}[!tp]
    \begin{minipage}[c]{.8\textwidth}
        \centering
        \subfigure[Clippy Warnings]{
        \label{fig:clippy_warning}
        \includegraphics[width=1\textwidth]{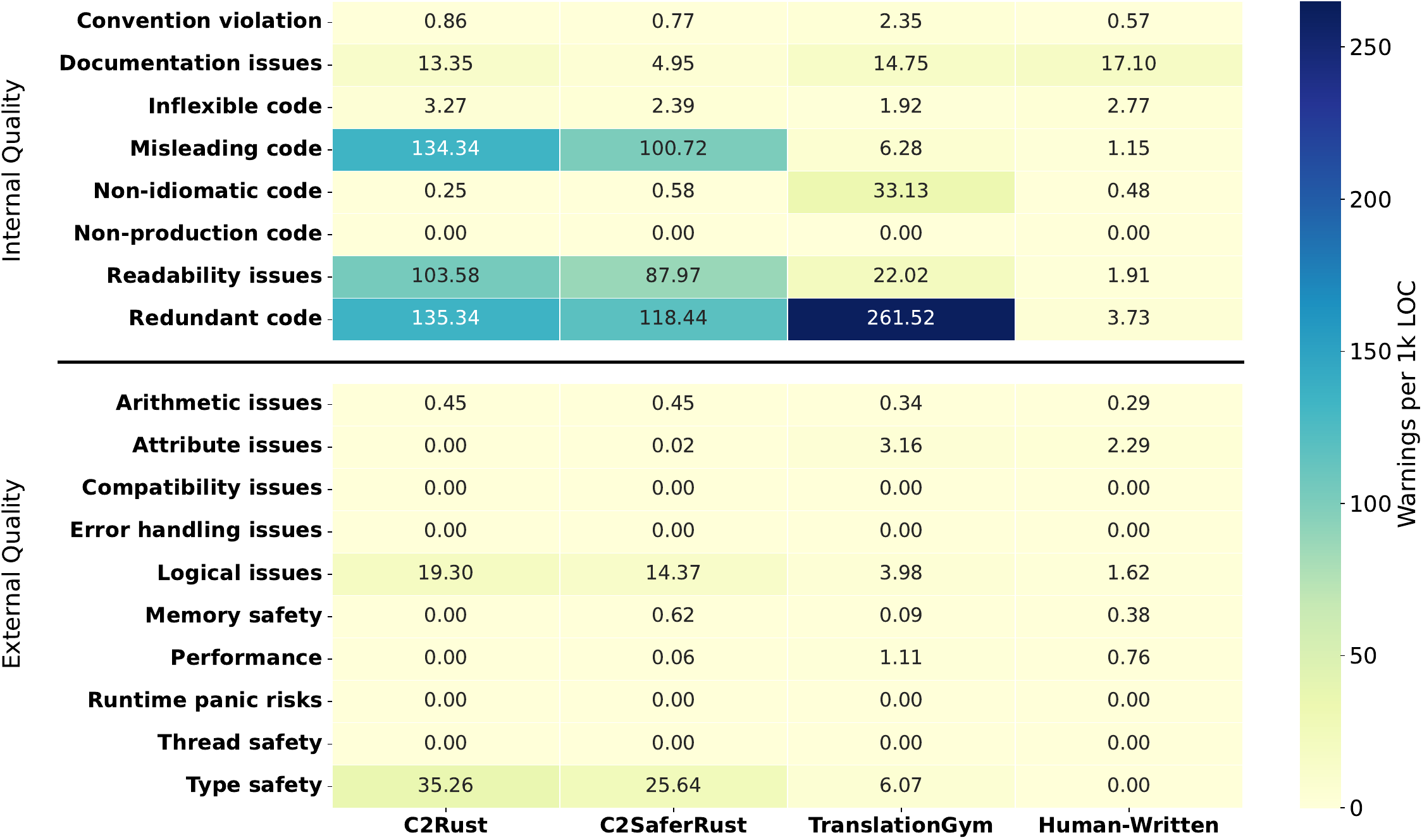}
        }
    \end{minipage}\vspace{1mm}\\%
    \begin{minipage}[c]{.8\textwidth}
        \centering
        \subfigure[GPT-4o Warnings]{
        \label{fig:LLM_warnings}
        \includegraphics[width=1\textwidth]{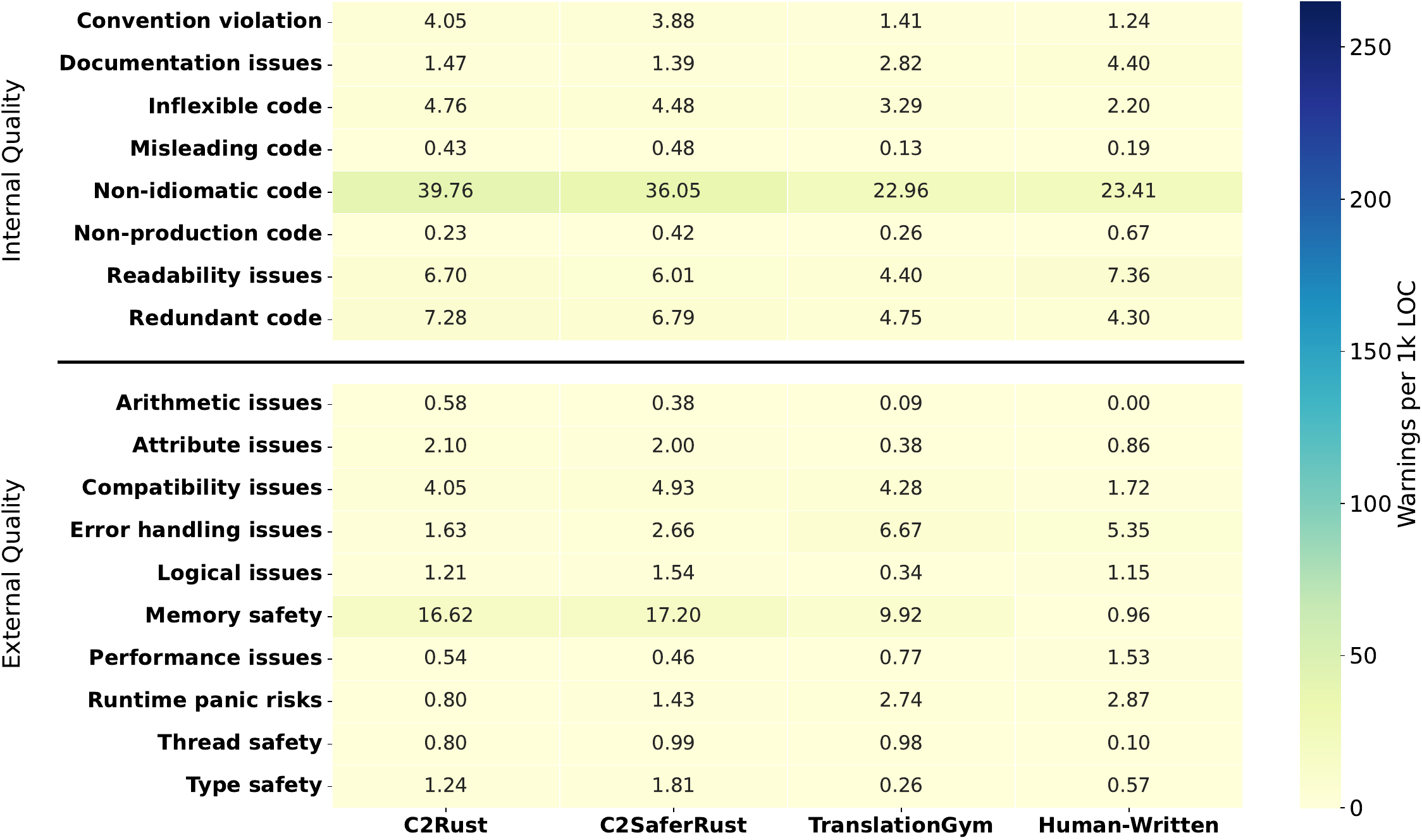}
        }
    \end{minipage}
    \caption{Number of Clippy and GPT-4o warnings, aggregated across all programs and normalized per 1k LOC.}
    \label{fig:warnings}    
\end{figure*}

Figure~\ref{fig:warnings} presents the total number of warnings, for both Clippy and GPT-4o, per issue category. The numbers are presented in normalized per 1k lines of code, since translations for each technique have different total lines of code (see Table~\ref{tab:programs}). Comparing raw warning counts would be misleading, as larger codebases would naturally report more warnings. Thus, normalization placed all techniques on the same scale. Raw warning counts are provided in the supplementary material~\cite{supplementary}.

\subsection{Comparison of Translation Techniques Across Quality Dimensions}

We can observe in Figure~\ref{fig:warnings} that each translation technique exhibits a distinct quality profile and that these profiles vary substantially depending on whether warnings are identified by a rule-based static analyzer (Clippy) or an LLM (GPT-4o).

Under Clippy, presented in Figure~\ref{fig:clippy_warning}, C2Rust and C2SaferRust are characterized by a heavy concentration of internal-quality issues. In particular, \textit{Misleading code}, \textit{Readability issues}, and \textit{Redundant code} dominate their warning profiles, collectively accounting for the majority of warnings per 1k LOC. This pattern is consistent with mechanically translated Rust that preserves C-like structure and control flow, leading to verbose, repetitive, and stylistically unidiomatic code. While C2SaferRust modestly reduces some categories compared to C2Rust, it does not fundamentally alter this quality signature, suggesting that safety-oriented rewrites alone are insufficient to improve maintainability-oriented dimensions captured by Clippy.
TranslationGym, by contrast, exhibits a different Clippy profile. This technique substantially eliminates \textit{Misleading code} and \textit{Readability issues} relative to both transpiler-based techniques (i.e., C2Rust and C2SaferRust), bringing these categories closer to the Human-Written baseline. However, this improvement comes at the cost of an extreme increase in Redundant code warnings, which are an order of magnitude higher than for any other technique. This result indicates that TranslationGym tends to replace low-level idiomatic violations with structurally repetitive or boilerplate-heavy code, likely reflecting localized rewrites that fail to consolidate common patterns across functions or modules. Thus, while TranslationGym improves certain surface-level Rust idioms, it introduces maintainability challenges of a different kind.
Human-Written translations achieve the lowest Clippy warning rates in most internal-quality categories, particularly \textit{Misleading code}, \textit{Readability issues}, and \textit{Redundant code}, reinforcing their role as a practical upper bound for idiomatic Rust. Nonetheless, they are not free of issues: \textit{Documentation} warnings remain non-trivial, making clear that expert translations prioritize functional correctness over exhaustive documentation.

We also observe in Figure~\ref{fig:clippy_warning} that several external quality categories, such as \textit{Compatibility issues}, \textit{Error handling}, \textit{Runtime panic risks}, and \textit{Thread safety}, have zero occurrences across all translation techniques. This also happens for \textit{Non-production code} in internal quality.
In a deeper analysis, we figure out that Clippy either (i) reported the underlying problems under different labels or (ii) did not emit any lint at all. 
For the former, we have an example in Listing~\ref{lst:C2RustB}. If this code is used in a multithreaded context, it could lead to data races. When the quality assessment tools evaluated this code, we had the following results: Clippy returned \textit{``Unsafe function's docs miss `\# Safety' section''} and GPT-4o returned \textit{``The function quote\_n\_mem takes a mutable reference to a global static variable, which is not thread-safe.''}  
Considering the message given by Clippy, this warning was grouped under ``\textit{Documentation issue}'', instead of appearing under ``\textit{Thread safety}''. Conversely, GPT-4o returned a better warning message, enabling proper classification.
For the latter case, the \CRust\ translation in Listing~\ref{lst:C2RustA} declares multiple \texttt{static mut} variables, which are globally mutable and unsafe to access concurrently. This allows unsynchronized access in multi-threaded contexts. Clippy raised no warning, and the GPT-4o returned \textit{``Using `static mut' variables without synchronization is unsafe in multi-threaded contexts. Rust would typically use thread-safe alternatives like `Mutex' or `AtomicPtr'.''} Thus, GPT-4o correctly flagged the issue as a ``Thread safety'' risk.
Another example is in Listing~\ref{lst:C2Rust7}, specifically for C2Rust, in which a raw unchecked pointer is stored in a global \texttt{static mut}, which clearly risks invalid memory access. However, Clippy did not flag this code, whereas GPT-4o correctly identified the risk. This indicates that Clippy primarily flagged issues in idiomatic Rust constructs while struggling to detect warnings in non-idiomatic translations.

\noindent 
\begin{minipage}[t]{0.48\textwidth} 
    \begin{lstlisting}[caption=Caught by Clippy but in a different category. Caught by GPT-4o correctly., captionpos=b, label=lst:C2RustB, language=Rust, frame=none, style=colouredRust, basicstyle=\footnotesize \ttfamily, numberstyle=\footnotesize \ttfamily]
pub static mut quote_quoting_options: quoting_options = {
    // ...
};
pub static mut quote_quoting_options: quoting_options = {
    // ...
};
pub unsafe extern "C" fn quote_n_mem(
    mut n: libc::c_int,
    mut arg: *const libc::c_char,
    mut argsize: size_t,
) -> *const libc::c_char {
    return quotearg_n_options(n, arg, argsize, &mut quote_quoting_options);
}	
\end{lstlisting}
\end{minipage}%
\hfill 
\begin{minipage}[t]{0.48\textwidth} 
    \begin{lstlisting}[caption=Issue uncaught by Clippy. Issue correctly caught by GPT-4o., captionpos=b, label=lst:C2RustA, language=Rust, frame=nones, style=colouredRust, basicstyle=\footnotesize \ttfamily, numberstyle=\footnotesize \ttfamily]
static mut stdout: *mut FILE;
static mut stderr: *mut FILE;
static mut optind: libc::c_int;
static mut Version: *const libc::c_char;
\end{lstlisting}
\end{minipage}

\noindent 
\begin{minipage}[t]{1\textwidth} 
    \centering
    \begin{lstlisting}[language=Rust, frame=nones, style=colouredRust, basicstyle=\footnotesize \ttfamily, numberstyle=\footnotesize \ttfamily, captionpos=b, label=lst:C2Rust7, caption={C2Rust issue uncaught by Clippy and Caught by GPT-4o.}]
static mut file_name: *const libc::c_char = 0 as *const libc::c_char;
#[no_mangle]
pub unsafe extern "C" fn close_stdout_set_file_name(mut file: *const libc::c_char) {
    file_name = file;
}
\end{lstlisting}
\end{minipage}

The comparison among techniques becomes more nuanced when examining GPT-4o's warnings, shown in Figure~\ref{fig:LLM_warnings}. GPT-4o identifies substantial numbers of issues in external-quality categories such as \textit{Compatibility issues}, \textit{Error handling}, \textit{Runtime panic risks}, and \textit{Thread safety} across all techniques. As discussed above, these categories are entirely absent in Clippy's results, revealing a key blind spot of rule-based analysis when applied to translated Rust code. With GPT-4o, C2Rust and C2SaferRust exhibit the highest levels of \textit{Memory safety} and \textit{Non-idiomatic code} concerns, while TranslationGym reduces these categories but still lags behind Human-Written code. Importantly, even Human-Written translations exhibit non-negligible \textit{Error handling} and \textit{Runtime panic risk} warnings, indicating that such issues persist despite careful manual translation.

\begin{tcolorbox}[colback=white, colframe=black, rounded corners=southwest, rounded corners=northwest, boxrule=0.5pt,
arc=8pt, left=2pt, right=2pt, top=2pt,bottom=2pt, fonttitle=\bfseries,before skip=5pt,after skip=5pt,breakable]
    \textbf{Finding.} \textit{Across techniques, a consistent trend emerges: improvements in one set of quality dimensions are often offset by regressions in others. TranslationGym moves closer to Human-Written code in stylistic and idiomatic dimensions but introduces substantial redundancy and does not fully eliminate robustness-related risks. C2Rust-style tools produce predictable, analyzable patterns but suffer from widespread maintainability issues. Finally, the divergence between Clippy and GPT-4o underscores that tool choice strongly influences which quality issues are visible: Clippy emphasizes syntactic and idiomatic concerns, while GPT-4o exposes semantic and behavioral risks that remain largely invisible to traditional linting. Together, these results demonstrate that no single translation technique uniformly dominates across quality dimensions and that a comprehensive assessment of translation quality requires complementary analysis techniques.}
\end{tcolorbox}

\begin{table}[!bp]
    \caption{Pairwise p-values from the statistical comparison of the Clippy and LLM warnings across translation techniques. Statistically significant (p-value < 0.05) and moderated (p-value < 0.15) are emphasized.}
    \footnotesize
    \label{tab:Statistical tests}
    \centering
    \begin{tabular}{l|c|c}
    \hline
        \textbf{Techniques} & \textbf{Clippy P-value} & \textbf{LLM P-value} \\
        \hline
        \CRust\ vs \CSaferRust\ & 0.347174 & 0.925379\\
        \CRust\ vs \TranslationGym & \textit{0.103343} & \textbf{0.035843} \\
        \CRust\ vs \HumanWritten & \textbf{0.000486} & \textbf{0.001114} \\
        \CSaferRust\ vs \TranslationGym & 0.925379 & 0.162904 \\
        \CSaferRust\ vs \HumanWritten & \textit{0.103343} & \textbf{0.010262} \\
        \TranslationGym\ vs \HumanWritten & 0.347174 & 0.728805 \\
    \hline
    \end{tabular}
\end{table}

To strengthen our analysis, we conducted a statistical test among the four translation techniques. The results in Table~\ref{tab:Statistical tests} indicate an overall statistically significant difference among the translation techniques for both Clippy (p-value=0.001213) and GPT-4o (p-value=0.000389). 
In the pair-wise comparison, both assessments show that \HumanWritten\ code is significantly better than \CRust\ (Clippy p-value=0.000486 and GPT-4o p-value=0.001114). 
This reflects the substantial structural and idiomatic refinements that human developers introduce compared to the mechanical ones of \CRust. For \CRust\ vs \TranslationGym, GPT-4o analysis indicates a significant difference (p-value=0.035843), while Clippy reports only a moderate effect (p-value=0.103343). This result suggests that while \TranslationGym\ introduces structural and logical improvements recognizable to the LLM, these changes do not fully register as stylistic or idiomatic refinements under Clippy’s rule-based lens. Similarly, \CSaferRust\ vs \HumanWritten\ shows meaningful differences in both tests, significant under GPT-4o analysis (p-value=0.010262) and moderate under Clippy (p-value=0.103343). This pattern shows that although \CSaferRust\ applies idiomatic patches over \CRust, it remains less aligned with human refinements in both style and deeper logical structure. In contrast, \CRust\ vs \CSaferRust\ yields consistent results across both analyses (Clippy p-value=0.347174 and GPT-4o p-value=0.925379), which shows that the safer variant largely inherits the baseline characteristics of \CRust\ without introducing substantial quality improvements. Finally, \TranslationGym\ vs \HumanWritten\ shows the closest alignment, with both analyses finding no significant difference (Clippy p-value=0.347174 and GPT-4o p-value=0.728805). This convergence indicates that \TranslationGym\ produces warning distributions more similar to \HumanWritten\ code.

\begin{tcolorbox}[colback=white, colframe=black, rounded corners=southwest, rounded corners=northwest, boxrule=0.5pt,
arc=8pt, left=2pt, right=2pt, top=2pt,bottom=2pt, fonttitle=\bfseries,before skip=5pt,after skip=5pt,breakable]
    \textbf{Finding.} \textit{Statistical tests confirm that \HumanWritten\ code is significantly better than \CRust. While \TranslationGym\ shows improvements over \CRust, recognizable by GPT-4o but less so by Clippy, \CSaferRust\ remains close to \CRust\ in quality. } 
\end{tcolorbox}

\subsection{Nature of Quality Issues in Translated Rust Code} 
\label{sec:qualitative_analysis}
We perform a deeper analysis to further investigate intriguing patterns identified in the results above and provide recommendations on how existing tools can be improved. 

\hypertarget{MemorySafetyExample}{\textbf{Memory Safety Observation.}}
In the translation of \texttt{dot\_or\_dotdot}, the \CRust\ version (Listing~\ref{lst:C2Rust1}) uses an \texttt{unsafe extern "C"} function and directly dereferences a raw pointer. The \CSaferRust\ version (Listing~\ref{lst:C2SaferRust1}) replaces this with a safer abstraction using \texttt{CStr::from\_ptr}, but removes the \texttt{unsafe} qualifier from the function signature.
Although the removal of \texttt{unsafe} in the \CSaferRust\ version was part of an attempt to clean up the code and make it more idiomatic using standard Rust abstractions like \texttt{CStr} and \texttt{\detokenize{to_string_lossy()}}, this introduces a subtler memory safety risk. The function still performs an unsafe operation \texttt{\detokenize{(CStr::from_ptr)}}, but without the \texttt{unsafe fn} boundary, this risk is no longer visible to the caller, violating Rust's safety principle. 

\noindent 
\begin{minipage}[t]{0.48\textwidth} 
    \begin{lstlisting}[caption=C2Rust version., captionpos=b, label=lst:C2Rust1, language=Rust, frame=nones, style=colouredRust, basicstyle=\footnotesize \ttfamily, numberstyle=\footnotesize \ttfamily]
#[inline]
unsafe extern "C" fn dot_or_dotdot(mut file_name: *const libc::c_char) -> bool {
    if *file_name.offset(0) as i32 == '.' as i32 {
        let sep: libc::c_char = *file_name.offset(
            ((*file_name.offset(1) as i32 == '.' as i32) as i32 + 1) as isize,
        );
        return sep == 0 || sep as i32 == '/' as i32;
    } else {
        return false;
    }
}
\end{lstlisting}
\end{minipage}%
\hfill 
\begin{minipage}[t]{0.48\textwidth} 
    \begin{lstlisting}[caption=C2SaferRust version, captionpos=b, label=lst:C2SaferRust1, language=Rust, frame=none, style=colouredRust, basicstyle=\footnotesize \ttfamily, numberstyle=\footnotesize \ttfamily]
#[inline]
fn dot_or_dotdot(file_name: *const libc::c_char) -> bool {
    let c_str = unsafe { std::ffi::CStr::from_ptr(file_name) };
    let file_name_str = c_str.to_string_lossy();
    if file_name_str.starts_with('.') {
        let sep = file_name_str.chars().nth(1);
        return sep.is_none() || sep.unwrap() == '/';
    } else {
        return false;
    }
}
\end{lstlisting}
\end{minipage}

\begin{tcolorbox}[colback=white, colframe=black, rounded corners=southwest, rounded corners=northwest, boxrule=0.5pt,
arc=8pt, left=2pt, right=2pt, top=2pt,bottom=2pt, fonttitle=\bfseries,before skip=5pt,after skip=5pt,breakable]
    \textbf{Recommendation 1.} \textit{ Preserve explicit \texttt{unsafe} boundaries when wrapping raw pointer operations.
    } 
    
    \textbf{Implication.}
    Translation tools aiming for safer or more idiomatic Rust should avoid superficial cleanups that remove \texttt{unsafe} without eliminating the underlying risk. Tool designers should:
    \begin{itemize}
        \item Maintain \texttt{unsafe fn} declarations unless the operation is fully encapsulated.
        \item Add safety comments when abstracting over unsafe logic.
    \end{itemize}
\end{tcolorbox}

\textbf{Runtime Panic Risks Observation.} 
In the example below, the \TranslationGym\ version (Listing~\ref{lst:TranslationGym2}) of \texttt{xrealloc\_rust} explicitly calls \texttt{panic!} when \texttt{new\_size} is zero or when the allocation fails. While this mirrors how some C programs abort on allocation failure, it introduces a non-recoverable panic into the Rust application. 
In contrast, the \CSaferRust\ version (Listing~\ref{lst:C2SaferRust2}) handles the null pointer and zero-size allocation case by returning a null pointer \texttt{(std::ptr::null\_mut())}, preserving compatibility with traditional C behavior and avoiding panics altogether.

\noindent 
\begin{minipage}[t]{0.54\textwidth} 
    \begin{lstlisting}[caption=TranslationGym version, captionpos=b, label=lst:TranslationGym2, language=Rust, frame=none, style=colouredRust, basicstyle=\footnotesize \ttfamily, numberstyle=\footnotesize \ttfamily]
fn xrealloc_rust<T>(ptr: Option<Box<T>>, new_size: usize) -> Box<T> {
    // If the size is 0, we should return None (equivalent to NULL in C)
    if new_size == 0 {
        panic!("Memory allocation failed");
    }
    unsafe {
        let ptr = alloc::alloc(layout);
        if ptr.is_null() {
             panic!("Memory allocation failed");
         }
   }
}
\end{lstlisting}
\end{minipage}%
\hfill 
\begin{minipage}[t]{0.42\textwidth} 
    \begin{lstlisting}[caption=C2SaferRust version., captionpos=b, label=lst:C2SaferRust2, language=Rust, frame=nones, style=colouredRust, basicstyle=\footnotesize \ttfamily, numberstyle=\footnotesize \ttfamily]
#[no_mangle]
pub fn xrealloc(p: *mut libc::c_void, s: usize) -> *mut libc::c_void {
    if p.is_null() && s == 0 {
        return std::ptr::null_mut();
    }
}
\end{lstlisting}
\end{minipage}

\begin{lstlisting}[caption=C2SaferRust version., captionpos=b, label=lst:C2SaferRust3, language=Rust, frame=none, style=colouredRust, basicstyle=\footnotesize \ttfamily, numberstyle=\footnotesize \ttfamily]
pub fn set_program_name(argv0: *const libc::c_char) {
    let c_str = unsafe { std::ffi::CStr::from_ptr(argv0) };
    let argv0_str = c_str.to_string_lossy();
    let slash = argv0_str.rfind('/').map(|index| &argv0_str[index + 1..]).unwrap_or(&argv0_str);
   if slash.len() >= 7 && slash.ends_with("/.libs/") {
        let base = &slash[7..];
        if base.starts_with("lt-") {
            unsafe {
                program_invocation_short_name = base[3..].as_ptr() as *mut libc::c_char;
            }
        }
        unsafe {
            program_name = base.as_ptr() as *mut libc::c_char;
            program_invocation_name = base.as_ptr() as *mut libc::c_char;
        }
    } else {
        unsafe {
            program_name = argv0 as *mut libc::c_char;
            program_invocation_name = argv0 as *mut libc::c_char;
        }
    }
}
\end{lstlisting}

\begin{lstlisting}[caption=C2Rust version.,
captionpos=b, label=lst:C2Rust3, language=Rust, frame=nones, style=colouredRust, basicstyle=\footnotesize \ttfamily, numberstyle=\footnotesize \ttfamily,
  floatplacement=!tp,]
pub unsafe extern "C" fn set_program_name(mut argv0: *const libc::c_char) {
    let mut slash: *const libc::c_char = 0 as *const libc::c_char;
    let mut base: *const libc::c_char = 0 as *const libc::c_char;
    slash = strrchr(argv0, '/' as i32);
    base = if !slash.is_null() {
        slash.offset(1 as libc::c_int as isize) }
    else {
        argv0
    };
    if base.offset_from(argv0) as libc::c_long >= 7 as libc::c_int as libc::c_long
    && strncmp(
        base.offset(-(7 as libc::c_int as isize)),
        b"/.libs/\0" as *const u8 as *const libc::c_char,
        7 as libc::c_int as libc::c_ulong,
    ) == 0 as libc::c_int {
        argv0 = base;
        if strncmp(
            base,
            b"lt-\0" as *const u8 as *const libc::c_char,
            3 as libc::c_int as libc::c_ulong,
        ) == 0 as libc::c_int {
            argv0 = base.offset(3 as libc::c_int as isize);
            program_invocation_short_name = argv0 as *mut libc::c_char;
        }
    }
    program_name = argv0;
    program_invocation_name = argv0 as *mut libc::c_char;
}
\end{lstlisting}

While the \TranslationGym\ implementation takes an idiomatic Rust approach by using \texttt{panic!} to signal allocation errors, this introduces runtime panic risks that immediately terminate the program. In high-reliability or FFI-heavy applications, such abrupt termination can lead to resource leaks or incomplete operations.  \CSaferRust\ avoids these risks by returning a null pointer, which, although not fully idiomatic in Rust, allows the calling code to handle the failure without crashing.

\begin{tcolorbox}[colback=white, colframe=black, rounded corners=southwest, rounded corners=northwest, boxrule=0.5pt,
arc=8pt, left=2pt, right=2pt, top=2pt,bottom=2pt, fonttitle=\bfseries,before skip=5pt,after skip=5pt,breakable]
    \textbf{Recommendation 2.} \textit{Avoid panicking on allocation failure, but prefer returning Result.} 
    
    \textbf{Implication.} Translation tools should:
    \begin{itemize}
        \item Replace \texttt{panic!} with error returning mechanisms (\texttt{Result}, \texttt{Option}) to allow for recovery.
        \item Avoid introducing non-recoverable runtime panic risks in memory allocation routines.
    \end{itemize}
\end{tcolorbox}

\textbf{Non-idiomatic Observation 1.}
While \CSaferRust\ (Listing~\ref{lst:C2SaferRust3}) eliminates unsafe pointer offset calculations found in the \CRust\ version  (Listing~\ref{lst:C2Rust3}), it retains C-like string processing patterns by manually searching for and slicing substrings. This approach can lead to subtle bugs in path handling, especially when dealing with platform-specific path separators or edge cases like relative paths, multiple slashes, or Unicode normalization. 
Rust’s \texttt{std::path::Path} and related utilities provide a more robust and platform-aware abstraction for file path operations. In contrast, the \CRust\ version’s direct pointer arithmetic is unsafe but directly mirrors the original C implementation’s low-level logic, which makes its operations explicit, although more error-prone.

\begin{tcolorbox}[colback=white, colframe=black, rounded corners=southwest, rounded corners=northwest, boxrule=0.5pt,
arc=8pt, left=2pt, right=2pt, top=2pt,bottom=2pt, fonttitle=\bfseries,before skip=5pt,after skip=5pt,breakable]
    \textbf{Recommendation 3.} \textit{Avoid manual string slicing for path manipulation. Use Rust's \texttt{std::path::Path} and APIs to perform platform-aware, idiomatic, and safer path operations.} 
    
    \textbf{Implication.} Translation tools should:
    \begin{itemize}
        \item Replace manual string slicing for path handling with idiomatic Rust abstractions such as \lstinline{std::path::Path} and \lstinline{PathBuf}.
        \item Ensure that translated code is robust to platform specific differences in path separators and normalization rules.
        \item Preserve clarity of intent by avoiding hidden assumptions in string operations and documenting transformations when low-level logic is replaced.
        \item Include automated checks to detect manual path parsing and suggest standard library replacements.
    \end{itemize}
\end{tcolorbox}

\textbf{Non-idiomatic Observation 2.}
\TranslationGym\ code (Listing~\ref{lst:TranslationGym4}) assumes that the target file is seekable and performs operations directly via \lstinline{lseek(...)}. It relies on detecting seek failures at runtime instead of preemptively verifying seekability. In contrast, \HumanWritten\  (Listing~\ref{lst:Human-Written4}) introduces a dedicated \lstinline{is_seekable} function that proactively checks whether the file supports seeking before performing any operations.

\noindent 
\begin{minipage}[t]{0.48\textwidth} 
    \begin{lstlisting}[caption=TranslationGym version., captionpos=b, label=lst:TranslationGym4, language=Rust, frame=nones, style=colouredRust, basicstyle=\footnotesize \ttfamily, numberstyle=\footnotesize \ttfamily,]
fn elseek_rust(fd: RawFd, offset: i64, whence: i32, filename: &str) -> i64 {
    // Use the libc lseek function directly since we need to maintain the file descriptor
    let new_offset = unsafe { libc::lseek(fd, offset, whence) };
    if new_offset < 0 {
        // Handle error case
        let error_msg = if whence == 0 {
            format!("{}: cannot seek to offset {}", filename, offset)
        } else {
            format!("{}: cannot seek to relative offset {}", filename, offset)
        };
    }
    new_offset
}
\end{lstlisting}
\end{minipage}%
\hfill 
\begin{minipage}[t]{0.48\textwidth} 
    \begin{lstlisting}[caption=Human-Written version., captionpos=b, label=lst:Human-Written4, language=Rust, frame=none, style=colouredRust, basicstyle=\footnotesize \ttfamily, numberstyle=\footnotesize \ttfamily]
fn is_seekable(input: &mut File) -> bool {
    let current_pos = input.stream_position();
    current_pos.is_ok()
        && input.seek(SeekFrom::End(0)).is_ok()
        && input.seek(SeekFrom::Start(current_pos.unwrap())).is_ok()
        // ... code continues
}
\end{lstlisting}
\end{minipage}

\TranslationGym\ reduces upfront checks and simplifies control flow, but at the cost of potentially invoking \lstinline{lseek} on non-seekable streams (e.g., pipes or sockets), which may incur unnecessary system call overhead and runtime errors. This design favors a ``fail and handle'' approach rather than explicit validation. Conversely, the \HumanWritten\ implementation ensures seekability through a proactive check, but does so using a non-idiomatic pattern, calling \lstinline{unwrap()} immediately after checking \lstinline{is_ok()}. While this prevents panics in most cases, it is verbose and deviates from Rust’s recommended error handling patterns, where the ``\lstinline{?}'' operator or pattern matching would make the intent clearer and reduce boilerplate.

\begin{tcolorbox}[colback=white, colframe=black, rounded corners=southwest, rounded corners=northwest, boxrule=0.5pt,
arc=8pt, left=2pt, right=2pt, top=2pt,bottom=2pt, fonttitle=\bfseries,before skip=5pt,after skip=5pt,breakable]
    \textbf{Recommendation 4.} \textit{Avoid assuming file capabilities such as seekability without verification. Use safe, high-level Rust APIs and idiomatic error handling (\texttt{?} operator or pattern matching) to validate capabilities before performing operations like \texttt{lseek}.} 
    
    \textbf{Implication.} Translation tools should:
    \begin{itemize}
        \item Detect operations that implicitly assume file capabilities (e.g., seekability) and either insert pre-checks or recommend their inclusion.
        \item Encourage the use of idiomatic Rust error handling mechanisms such as \lstinline{?} or pattern matching instead of redundant \lstinline{is_ok()} + \lstinline{unwrap()} patterns.
        \item Where performance is critical, provide configuration options for either pre-checking capabilities or relying on error-driven control flow.
        \item Offer automated suggestions for replacing direct system calls like \lstinline{lseek} with safe, high-level Rust APIs where possible.
    \end{itemize}
\end{tcolorbox}

\noindent 
\begin{minipage}[t]{0.48\textwidth} 
    \begin{lstlisting}[caption=C2Rust version., captionpos=b, label=lst:C2Rust5, language=Rust, frame=nones, style=colouredRust, basicstyle=\footnotesize \ttfamily, numberstyle=\footnotesize \ttfamily]
static mut infile: *const libc::c_char = 0 as *const libc::c_char;
unsafe extern "C" fn copy_cat() -> libc::c_int {
// ...
        error(
            0 as libc::c_int,
            *__errno_location(),
            b"%s\0" as *const u8 as *const libc::c_char,
            quotearg_n_style_colon(
                /*...*/,
                infile,
            ),
        );        
    // ...
  }
\end{lstlisting}
\end{minipage}%
\hfill 
\begin{minipage}[t]{0.48\textwidth} 
    \begin{lstlisting}[caption=C2SaferRust version., captionpos=b, label=lst:C2SafeRust5, language=Rust, frame=none, style=colouredRust, basicstyle=\footnotesize \ttfamily, numberstyle=\footnotesize \ttfamily]
unsafe extern "C" fn copy_cat() -> libc::c_int {
// ...
error(
        0,
        *__errno_location(),
        b"%s\0" as *const u8 as *const libc::c_char,
        quotearg_n_style_colon(
        /*...*/,
        infile, // again, no wrapper
        ),
);
}
\end{lstlisting}
\end{minipage}
\\%
\begin{minipage}[t]{1\textwidth} 
\begin{lstlisting}[caption=TranslationGym version., captionpos=b, label=lst:TranslationGym5, language=Rust, frame=nones, style=colouredRust, basicstyle=\footnotesize \ttfamily, numberstyle=\footnotesize \ttfamily]
impl InfileWrapper {
    pub fn new() -> Self {
        let val = Self::read_global();
        InfileWrapper { val }
    }
    pub fn get(&self) -> Option<String> {
        Self::read_global()
    }
    // ...
}
fn copy_cat_rust(infile_wrapper: &InfileWrapper) -> i32 {
    // ...
    if let Some(filename) = infile_wrapper.get() {
        eprintln!("{}: {}", quotearg_n_style_colon_rust(0, shell_escape_quoting_style(), &filename));
    }
    return -1;
}
\end{lstlisting}
\end{minipage}

\textbf{Convention violations.}
Both \CRust\ (Listing~\ref{lst:C2Rust5}) and \CSaferRust\ (Listing~\ref{lst:C2SafeRust5}) rely on a global mutable static pointer \texttt{infile} to track the file name, which is unsafe, straightforward, and typical of C-style code. The \TranslationGym\ version (Listing~\ref{lst:TranslationGym5}) replaces this with a safer wrapper type, \texttt{InfileWrapper}, encapsulating access to the global value. This improves safety and abstraction by avoiding direct manipulation of mutable static pointers. However, Clippy flags the absence of a \texttt{Default} implementation for \texttt{InfileWrapper} as non-idiomatic. In Rust, wrapper types that can be easily and meaningfully initialized, such as via \texttt{InfileWrapper::new()}, are expected to implement \texttt{Default}, which enables ergonomic usage like \texttt{InfileWrapper::default()} or struct field initialization with \texttt{..Default::default()}. 

\TranslationGym\ replaces a global mutable pointer with a dedicated wrapper type, improving safety by encapsulating state and controlling access. However, omitting a \texttt{Default} implementation reduces the ease of initialization in contexts where a default value is desired. Implementing \texttt{Default} would make the wrapper more flexible and idiomatic, and align it with Rust’s conventions for simple, self-contained types.
\begin{tcolorbox}[colback=white, colframe=black, rounded corners=southwest, rounded corners=northwest, boxrule=0.5pt,
arc=8pt, left=2pt, right=2pt, top=2pt,bottom=2pt, fonttitle=\bfseries,before skip=5pt,after skip=5pt,breakable]
    \textbf{Recommendation 5.} \textit{Implement the \texttt{Default} trait for simple wrapper types that have a clear, sensible initialization.}

    \textbf{Implication.}
    Translation tools should:  
\begin{itemize}
    \item Replace global mutable pointers with wrapper types to improve safety and encapsulation.  
    \item Implement \texttt{Default} for wrapper types when a natural initialization exists.  
    \item Ensure generated code follows idiomatic Rust patterns to reduce friction for developers integrating translated components. 
\end{itemize}
\end{tcolorbox}

\textbf{Documentation issues.}
The \HumanWritten\ code often prioritizes completeness over precision, including detailed documentation that conveys design intent and technical context. However, this results in a Clippy warning despite its higher informational value.
In the \texttt{write\_fast\_using\_splice} function (Listing~\ref{lst:HumanWritten8}), the doc comment explains the use of \texttt{splice()} and its performance benefits in detail. While this information is valuable, Clippy flags it because the first paragraph in Rust documentation is expected to be a concise, one-sentence summary. This convention improves readability and compatibility with auto-generated documentation tools such as \texttt{rustdoc}. In contrast, the \TranslationGym\ example (Listing~\ref{lst:TranslationGym6}) provides minimal but structurally compliant comments, which align with the stylistic requirements and pass without warnings.

    \begin{lstlisting}[caption={Human-Written version.}, captionpos=b, label=lst:HumanWritten8, language=Rust, frame=nones, style=colouredRust, basicstyle=\footnotesize \ttfamily, numberstyle=\footnotesize \ttfamily]
/// This function is called from `write_fast()` on Linux and Android. The
/// function `splice()` is used to move data between two file descriptors
/// without copying between kernel and user spaces. This results in a large
/// speedup.
///
/// The `bool` in the result value indicates if we need to fall back to normal
/// copying or not. False means we don't have to.
#[inline]
pub(super) fn write_fast_using_splice<R: FdReadable, S: AsRawFd + AsFd>(){}
\end{lstlisting}

    \begin{lstlisting}[caption=TranslationGym version., captionpos=b, label=lst:TranslationGym6, language=Rust, frame=nones, style=colouredRust, basicstyle=\footnotesize \ttfamily, numberstyle=\footnotesize \ttfamily]
/// Attempts to write the entire buffer to the given file descriptor.
/// Returns the total number of bytes written.
fn full_write_rust(fd: RawFd, buf: &[u8]) -> usize {}
\end{lstlisting}


\begin{tcolorbox}[colback=white, colframe=black, rounded corners=southwest, rounded corners=northwest, boxrule=0.5pt,
arc=8pt, left=2pt, right=2pt, top=2pt,bottom=2pt, fonttitle=\bfseries,before skip=5pt,after skip=5pt,breakable]
    \textbf{Recommendation 6.} \textit{Translation techniques should follow Rust documentation conventions. Rust’s documentation guidelines~\cite{RustDoc2025} emphasize that the opening paragraph of a doc comment should serve as a brief summary. Longer technical explanations should be placed in subsequent paragraphs to preserve clarity in generated documentation and to provide quick scanning readers with an immediate understanding of the function’s purpose.}
    
    \textbf{Implication.}
     Tool designers and developers should:
\begin{itemize}
    \item Keep the first paragraph of documentation comments concise and to the point.
    \item Move detailed technical explanations, performance notes, and design rationale into subsequent paragraphs.
    \item Balance informativeness with adherence to Rust’s idiomatic documentation style to maximize both clarity and lint compliance.
\end{itemize}
\end{tcolorbox}

\subsection{Discussion}

The results show a progressive improvement in quality from \CRust\ to \HumanWritten, but with trade-offs at each stage. Across both Clippy and GPT-4o, we observed that when translation tools attempt to improve on \CRust\ by introducing safer abstractions or more idiomatic constructs, they often introduce new problems, such as runtime panic risks, performance costs, or hidden safety concerns. Similarly, categories like readability, documentation, and non-idiomatic usage persist regardless of the technique, which shows that these challenges are not easily addressed by either automated or manual effort. These consistent patterns across these techniques show that translation quality is not a simple matter of reducing warnings in one dimension, but of balancing competing aspects of internal and external quality.

From these findings, we see that producing high-quality Rust translations requires more than simply converting C syntax into Rust syntax. Effective translation demands attention to deeper aspects of code quality, such as safety, idiomatic style, performance, and maintainability. For example, our analysis points to the importance of preserving explicit unsafe boundaries, avoiding panics in error handling, and adopting standard Rust APIs for memory and path management. These are only a few of the improvements we identified; other categories of issues also reveal opportunities for refinement. Importantly, even \HumanWritten\ translations still face challenges in categories such as readability, documentation, and non-idiomatic code, among others. This shows that producing high-quality Rust code involves complexities that persist beyond automated translation, and points to the need for stronger tool support and clearer guidance. Analyzing results through these quality categories allowed us to move beyond aggregate warning counts and reason about why specific translation strategies succeed or fail. This category-driven analysis exposed recurring trade-offs across techniques and directly informed the concrete recommendations presented in the paper. At the same time, it offers a structured way to reason about code quality that can be used for future studies of translation tools and Rust code, especially given that existing translation work focuses almost exclusively on correctness and unsafe usage.

Our analysis highlights the robustness and limitations of current code quality evaluation tools, particularly Clippy and GPT-4, when applied to Rust code from C. Clippy, widely recognized as a leading static analyzer for Rust, demonstrates strong proficiency in detecting unsafe code patterns and enforcing idiomatic Rust standards. However, our results clearly show that Clippy's effectiveness significantly diminishes when analyzing C-style Rust code. For instance, Clippy includes lints like \lstinline{uninit_vec}, explicitly designed to flag uninitialized usages of standard Rust constructs such as Vec. These lints naturally fail to trigger in code that is heavily reliant on raw pointers and manual memory management, typical of translations from C, because it does not use idiomatic Rust abstractions that Clippy expects. Consequently, critical memory-safety issues and subtle risks in unsafe code blocks often escape Clippy's detection, revealing a significant blind spot in its current design when evaluating transpiled or pointer-heavy Rust code.

Using  GPT-4 as a linter revealed both strengths and notable weaknesses. Unlike Clippy, GPT-4 shows considerable flexibility and depth, effectively identifying nuanced memory-safety concerns, non-idiomatic patterns, and unsafe constructs, even in complex, pointer-heavy Rust translations. However, GPT-4 introduced inconsistencies and inaccuracies in its categorization of warnings. Despite clear instructions, GPT-4 occasionally generated new, unintended warning categories not included in our predefined set of 18 categories. Specifically, we observed unexpected categorizations such as \lstinline{magic_numbers} (6 instances), inconsistency (8 instances), \lstinline{undefined_symbols} (3 instances), \lstinline{unused_imports} (2 instances), and isolated cases such as \lstinline{code_duplication}, \lstinline{unreachable_code}, \lstinline{inconsistent_style}, and \lstinline{missing_imports}. Although these extraneous categories were relatively few compared to correctly categorized warnings, their occurrence shows the unpredictable nature and extraneous categorizations occasionally produced by LLM-based analyses.

Overall, our findings show a complementary relationship between Clippy and GPT-4. While Clippy reliably enforces idiomatic Rust practices and catches straightforward unsafe patterns within idiomatic contexts, GPT-4 effectively fills the gaps left by Clippy in analyzing non-idiomatic and pointer-intensive code. Nevertheless, reliance on GPT-4 alone could pose challenges due to its less-deterministic behavior and occasional unpredictability in its categorized warnings.

\subsection{Threats to Validity}

In this section, we discuss the threats to the validity of our study, and how we mitigate them~\cite{Wohlin2000}.

\textbf{Construct Validity.} Code quality was evaluated through warnings produced by Clippy and GPT-4o, which could introduce inherent limitations. Clippy relies on rule-based lints that emphasize syntactic, idiomatic, and localized structural properties, potentially under-representing semantic and behavioral concerns. GPT-4o, in contrast, uses learned representations that may introduce subjectivity in issue identification. To mitigate these threats, we classify warnings into a shared taxonomy and analyze results at the category level, reducing sensitivity to tool-specific behaviors.

\textbf{Internal Validity.} Internal validity may be affected by factors beyond our direct control, as we relied on translated artifacts reported in prior work. Consequently, differences among techniques may reflect not only translation strategies but also implementation details used in the original studies. To mitigate this threat, we analyze all translated artifacts using a uniform evaluation pipeline, applying identical Clippy and GPT-4o analyses and normalizing warning counts per 1k LOC. While this approach ensures consistency in quality assessment, we cannot fully isolate the impact of translation techniques from upstream experimental choices made in the source studies.

\textbf{External Validity.} The generalizability of our findings is limited by the evaluated programs and techniques. The subject programs may not capture the full diversity of C codebases, and the selected techniques represent only a subset of available ones. However, the inclusion of three techniques and human-written translations increases the relevance of our results to both research and practice.

\textbf{Conclusion Validity.} Warnings are not equivalent to defects, and different categories vary in practical impact. We therefore avoid claims about correctness or safety guarantees and interpret warnings as relative indicators of potential quality risks. Qualitative analysis of representative issues further grounds our conclusions.

\section{Related Work}
\label{sec:related_work}

Recent research has devoted considerable attention to developing tools that automate or support the translation of C programs into Rust. Transpilers such as \CRust~\cite{C2Rust2018} convert C code into functionally equivalent Rust, but the resulting code is often non-idiomatic and extensively uses unsafe blocks, undermining Rust's safety guarantees. Other efforts have focused on techniques that take \CRust’s unsafe output and attempt to make it safer. For example, Laertes~\cite{Mehmet2023} uses compiler feedback to iteratively rewrite some raw pointers as borrows, though this only applies to a subset of unsafe cases and covers a limited fraction of functions. CROWN~\cite{Hanliang2023} derives ownership constraints for raw pointers and uses an SMT solver to replace them with references, but it is similarly restricted to certain pointer categories. 

More recently, \CSaferRust~\cite{C2SaferRust2025} has combined \CRust\ with LLM-based refinements to incrementally transform unsafe Rust into safer and more idiomatic Rust while preserving correctness through testing. Other approaches, such as \TranslationGym~\cite{TranslationGym2025}, bypass the transpilation step entirely and use LLMs for direct translation from C to Rust, using an iterative process of compilation and test feedback to repair the generated code. Along similar lines, AlphaTrans \cite{Ibrahimzada2025} proposes a neuro-symbolic, repository-level translation approach that decomposes large projects into smaller fragments and validates translations using existing tests to ensure functional correctness. While these tools show clear progress in enabling automated migration, they are primarily evaluated on their ability to reduce unsafe usage or maintain functional correctness, with little emphasis on broader aspects of code quality. Metrics such as number of raw pointers, unsafe lines of code, or the success rate of translations, which are used by those translation tools are important for assessing safety but overlook internal quality aspects like readability, maintainability, and documentation, as well as external quality factors such as clarity of error handling or robustness of control flow. 

Recent work by Dehghan et al.~\cite{dehghan2025translatinglargescalecrepositories} moves beyond correctness and safety by explicitly targeting idiomaticity and readability in C-to-Rust translation. These authors evaluate translated code using a diverse set of quality metrics and show that code quality can be measured more broadly, but they focus on designing and evaluating a single translation pipeline rather than comparing quality trade-offs across different translation strategies. Related work has also explored repository-level translation validation, repair, and bug characterization across multiple programming languages, but these studies primarily emphasize functional equivalence rather than comparative analysis of code quality dimensions~\cite{ke2025advancingautomatedinisolationvalidation,ibrahimzada2025matchfixagentlanguageagnosticautonomousrepositorylevel, Pan_2024}.

Despite prior work, little is known about trade-offs in quality that arise when different translation strategies are applied, particularly how improvements in one category may introduce issues in another. Existing evaluations highlight progress in reducing unsafe constructs, but they do not systematically examine how Rust-translated code compares across a wider set of quality dimensions. Our work addresses this gap by providing an empirical analysis of translated Rust code across both internal and external quality factors. 

\section{Conclusion}
\label{sec:conclusion}

In this paper, we conducted an empirical study on the quality of Rust-translated code from C. We compared three automated approaches, \CRust, \CSaferRust, and \TranslationGym\, as well as \HumanWritten\, translations by assessing them across 18 categories, including internal and external quality. Our evaluation relied on Clippy, a rule-based Rust linter, and GPT-4o, which provided complementary coverage of unsafe and non-idiomatic patterns.

Our study shows a progression in translation quality. While automated techniques like \CSaferRust\ and \TranslationGym\ build upon \CRust\ to improve external (e.g., memory safety) and internal quality (e.g., non-idiomatic code), these gains often come at the external quality's expense, introducing new risks in categories like compatibility issues, runtime panic risks, and thread safety. Notably, even \HumanWritten\ translations exhibit issues in all categories although relatively small compared to the other techniques. However, it shows more internal quality issues like poor documentation and readability, which overall shows Rust's inherent complexity. Therefore, we recommend tool builders to prioritize strategies that avoid trading one quality for the other and instead try to address issues from different aspects of code quality. 

Overall, this study contributes a taxonomy of translation-related Rust quality issues, a comparative analysis of automated and human translation strategies, and practical recommendations for tool builders and developers. Future work includes extending the study to larger systems and investigating hybrid approaches that combine static analysis with LLM-based reasoning to improve the reliability of translated Rust code.

\section*{Data Availability}
All source code, collected data, and complementary results are in the supplementary material~\cite{supplementary}.

\bibliographystyle{ACM-Reference-Format}
\bibliography{references}

@misc{supplementary,
    author={Anonymous},
    title={Supplementary material as part of the paper},
    year={2025},
    note = {\url{https://doi.org/10.5281/zenodo.17102148}}
}

@article{Emre2021,
author = {Emre, Mehmet and Schroeder, Ryan and Dewey, Kyle and Hardekopf, Ben},
title = {Translating C to safer Rust},
year = {2021},
issue_date = {October 2021},
publisher = {ACM},
address = {New York, NY, USA},
volume = {5},
number = {OOPSLA},
doi = {10.1145/3485498},
journal = {Proc. ACM Program. Lang.},
month = oct,
articleno = {121},
numpages = {29},
}

@article{Popescu2025,
author = {Popescu, Lucian and Lopes, Nuno P.},
title = {Exploiting Undefined Behavior in C/C++ Programs for Optimization: A Study on the Performance Impact},
year = {2025},
issue_date = {June 2025},
publisher = {Association for Computing Machinery},
address = {New York, NY, USA},
volume = {9},
number = {PLDI},
journal = {Proc. ACM Program. Lang.},
month = jun,
articleno = {161},
numpages = {24},
}

@online{TIOBE2026,
  author = {{The TIOBE Programming Community}},
  title = {TIOBE Index for January 2026},
  year = 2026,
  url = {https://www.tiobe.com/tiobe-index/},
  urldate = {visit in Jan 202}
}

@online{WhiteHouse2021,
  author = {{The White House}},
  title = {Executive Order on Improving the Nation’s Cybersecurity},
  year = 2021,
  url = {https://www.cisa.gov/topics/cybersecurity-best-practices/executive-order-improving-nations-cybersecurity},
  urldate = {visit in Jan 2026}
}

@INPROCEEDINGS{Pereira2022,
author={Pereira, José D'Abruzzo and Antunes, João Henggeler and Vieira, Marco},
booktitle={27th Pacific Rim International Symposium on Dependable Computing (PRDC)}, 
title={A Software Vulnerability Dataset of Large Open Source C/C++ Projects}, 
year={2022},
pages={152-163}
}

@inproceedings{durumeric2014,
  title={The matter of heartbleed},
  author={Durumeric, Zakir and Li, Frank and Kasten, James and Amann, Johanna and Beekman, Jethro and Payer, Mathias and Weaver, Nicolas and Adrian, David and Paxson, Vern and others},
  booktitle={Internet Measurement Conference},
  pages={475--488},
  year={2014}
}

@article{leveson1993,
  title={An investigation of the Therac-25 accidents},
  author={Leveson, Nancy G and Turner, Clark S},
  journal={Computer},
  volume={26},
  number={7},
  pages={18--41},
  year={1993},
  publisher={IEEE}
}

@incollection{shetty2019crust,
  title={Crust: AC/C++ to Rust transpiler using a “nano-parser methodology” to avoid C/C++ safety issues in legacy code},
  author={Shetty, Nishanth and Saldanha, Nikhil and Thippeswamy, MN},
  booktitle={Emerging Research in Computing, Information, Communication and Applications: ERCICA 2018, Volume 1},
  pages={241--250},
  year={2019},
  publisher={Springer}
}

@book{klabnik2023rust,
  title={The Rust programming language},
  author={Klabnik, Steve and Nichols, Carol},
  year={2023},
  publisher={No Starch Press}
}

@inproceedings{Li2025, 
   title={Translating C To Rust: Lessons from a User Study},
   booktitle={Network and Distributed System Security Symposium (NDSS)},
   publisher={Internet Society},
   author={Li, Ruishi and Wang, Bo and Li, Tianyu and Saxena, Prateek and Kundu, Ashish},
   year={2025}
}

@book{spinellis2006code,
  title={Code quality: the open source perspective},
  author={Spinellis, Diomidis},
  year={2006},
  publisher={Adobe Press}
}

@inproceedings{Simoes2024,
author = {Sim\~{o}es, Igor Regis da Silva and Venson, Elaine},
title = {Evaluating Source Code Quality with Large Language Models: a comparative study},
year = {2024},
isbn = {9798400717772},
publisher = {Association for Computing Machinery},
address = {New York, NY, USA},
doi = {10.1145/3701625.3701650},
booktitle = {XXIII Brazilian Symposium on Software Quality},
pages = {103–113}
}

@misc{yetistiren2023,
title={Evaluating the Code Quality of AI-Assisted Code Generation Tools: An Empirical Study on GitHub Copilot, Amazon CodeWhisperer, and ChatGPT}, 
author={Burak Yetiştiren and Işık Özsoy and Miray Ayerdem and Eray Tüzün},
year={2023},
eprint={2304.10778},
archivePrefix={arXiv},
primaryClass={cs.SE},
url={https://arxiv.org/abs/2304.10778}, 
}

@inproceedings{li2024clippy,
  title={Unleashing the power of clippy in real-world rust projects},
  author={Li, Chunmiao and Yu, Yijun and Wu, Haitao and Carlig, Luca and Nie, Shijie and Jiang, Lingxiao},
  booktitle={46th International Conference on Software Engineering: Companion Proceedings},
  pages={318--319},
  year={2024}
}

@misc{C2Rust2018,
title={C2Rust}, 
author={Galois},
year={2018},
url={https://www.galois.com/articles/c2rust}, 
}

@article{Mehmet2023,
author = {Emre, Mehmet and Boyland, Peter and Parekh, Aesha and Schroeder, Ryan and Dewey, Kyle and Hardekopf, Ben},
title = {Aliasing Limits on Translating C to Safe Rust},
year = {2023},
issue_date = {April 2023},
publisher = {Association for Computing Machinery},
address = {New York, NY, USA},
volume = {7},
number = {OOPSLA1},
journal = {Proc. ACM Program. Lang.},
month = apr,
articleno = {94},
numpages = {29}
}

@InProceedings{Hanliang2023,
author="Zhang, Hanliang
and David, Cristina
and Yu, Yijun
and Wang, Meng",
editor="Enea, Constantin
and Lal, Akash",
title="Ownership Guided C to Rust Translation",
booktitle="Computer Aided Verification",
year="2023",
publisher="Springer Nature Switzerland",
address="Cham",
pages="459--482",
isbn="978-3-031-37709-9"
}

@misc{C2SaferRust2025,
      title={C2SaferRust: Transforming C Projects into Safer Rust with NeuroSymbolic Techniques}, 
      author={Vikram Nitin and Rahul Krishna and Luiz Lemos do Valle and Baishakhi Ray},
      year={2025},
      eprint={2501.14257},
      archivePrefix={arXiv},
      primaryClass={cs.SE},
      url={https://arxiv.org/abs/2501.14257}, 
}

@misc{TranslationGym2025,
	author = {Vikram Nitin},
	title = {Translation Gym},
	url = {https://github.com/vikramnitin9/translation_gym},
	year = {2025},
}

@misc{uutils2025,
	author = {{uutils}},
	title = {Cross-platform Rust rewrite of the GNU coreutils },
	url = {https://github.com/uutils/coreutils},
	year = {2025},
}

@misc{ISO25010,
	author = {{International Organization for Standardization}},
	title = {Systems and software engineering — Systems and software Quality Requirements and Evaluation (SQuaRE) — Product quality model},
	url = {https://www.iso.org/standard/78176.html},
	year = {2023},
}

@misc{Thomas2023,
	author = {Thomas Claburn},
	month = {4},
	title = {Microsoft is busy rewriting core Windows code in memory-safe Rust},
	url = {https://www.theregister.com/2023/04/27/microsoft_windows_rust/},
	year = {2023}
}

@misc{coreutils2025,
	author = {{coreutils}},
	title = {GNU core utilities},
	url = {https://github.com/coreutils/coreutils},
	year = {2025},
}

@article{hong2025,
title={Automatically Translating C to Rust},
author={Hong, Jaemin and Ryu, Sukyoung},
journal={Communications of the ACM},
volume={68},
number={11},
pages={58--65},
year={2025},
publisher={ACM}
}

@Book{Wohlin2000,
  author    = {Wohlin, Claes and Runeson, Per and Höst, Martin and Ohlsson, Magnus C. and Regnell, Bjöorn and Wesslén, Anders},
  publisher = {Kluwer Academic Publishers},
  title     = {Experimentation in Software Engineering: An Introduction},
  year      = {2000},
  address   = {USA},
  isbn      = {0792386825},
  doi       = {10.1007/978-1-4615-4625-2},
}

@article{damasceno2023test,
  title={Test smell refactoring revisited: What can internal quality attributes and developers’ experience tell us?},
  author={Damasceno, Humberto and Bezerra, Carla and Campos, Denivan and Machado, Ivan and Coutinho, Emanuel},
  journal={Journal of Software Engineering Research and Development},
  volume={11},
  number={1},
  pages={13--1},
  year={2023}
}

@inproceedings{lee2022engineering,
  title={The Engineering Implications of Code Maintenance in Practice},
  author={Lee, Noah and Abreu, Rui and Yatbaz, Mehmet and Qu, Hang and Nagappan, Nachiappan},
  booktitle={International Conference on Software Maintenance and Evolution (ICSME)},
  pages={568--577},
  year={2022},
  organization={IEEE}
}

@inproceedings{cruzes2011,
  title={Recommended steps for thematic synthesis in software engineering},
  author={Cruzes, Daniela S and Dyba, Tore},
  booktitle={International Symposium on Empirical Software Engineering and Measurement (ESEM)},
  pages={275--284},
  year={2011},
  organization={IEEE}
}

@misc{GPT4o2024,
	author = {{OpenAPI}},
	title = {Hello GPT-4o},
	url = {https://openai.com/index/hello-gpt-4o/},
	year = {2024},
}

@misc{RustDoc2025,
	author = {{Rust Team}},
	title = {The rustdoc book: How to write documentation},
	url = {https://doc.rust-lang.org/rustdoc/how-to-write-documentation.html},
	year = {2025},
}

@article{friedman1937use,
  title={The use of ranks to avoid the assumption of normality implicit in the analysis of variance},
  author={Friedman, Milton},
  journal={Journal of the American Statistical Association},
  volume={32},
  number={200},
  pages={675--701},
  year={1937},
  publisher={Taylor \& Francis}
}

@book{nemenyi1963distribution,
  title={Distribution-free multiple comparisons.},
  author={Nemenyi, Peter Bjorn},
  year={1963},
  publisher={Princeton University}
}

@article{Ibrahimzada2025,
author = {Ibrahimzada, Ali Reza and Ke, Kaiyao and Pawagi, Mrigank and Abid, Muhammad Salman and Pan, Rangeet and Sinha, Saurabh and Jabbarvand, Reyhaneh},
title = {AlphaTrans: A Neuro-Symbolic Compositional Approach for Repository-Level Code Translation and Validation},
year = {2025},
issue_date = {July 2025},
publisher = {Association for Computing Machinery},
address = {New York, NY, USA},
volume = {2},
number = {FSE},
url = {https://doi.org/10.1145/3729379},
doi = {10.1145/3729379},
journal = {Proc. ACM Softw. Eng.},
month = jun,
articleno = {FSE109},
numpages = {23}
}

@misc{dehghan2025translatinglargescalecrepositories,
      title={Translating Large-Scale C Repositories to Idiomatic Rust}, 
      author={Saman Dehghan and Tianran Sun and Tianxiang Wu and Zihan Li and Reyhaneh Jabbarvand},
      year={2025},
      eprint={2511.20617},
      archivePrefix={arXiv},
      primaryClass={cs.SE},
      url={https://arxiv.org/abs/2511.20617}, 
}

@misc{ke2025advancingautomatedinisolationvalidation,
      title={Advancing Automated In-Isolation Validation in Repository-Level Code Translation}, 
      author={Kaiyao Ke and Ali Reza Ibrahimzada and Rangeet Pan and Saurabh Sinha and Reyhaneh Jabbarvand},
      year={2025},
      eprint={2511.21878},
      archivePrefix={arXiv},
      primaryClass={cs.SE},
      url={https://arxiv.org/abs/2511.21878}, 
}

@inproceedings{Pan_2024, series={ICSE ’24},
   title={Lost in Translation: A Study of Bugs Introduced by Large Language Models while Translating Code},
   url={http://dx.doi.org/10.1145/3597503.3639226},
   DOI={10.1145/3597503.3639226},
   booktitle={Proceedings of the IEEE/ACM 46th International Conference on Software Engineering},
   publisher={ACM},
   author={Pan, Rangeet and Ibrahimzada, Ali Reza and Krishna, Rahul and Sankar, Divya and Wassi, Lambert Pouguem and Merler, Michele and Sobolev, Boris and Pavuluri, Raju and Sinha, Saurabh and Jabbarvand, Reyhaneh},
   year={2024},
   month=apr, pages={1–13},
   collection={ICSE ’24} }

@misc{ibrahimzada2025matchfixagentlanguageagnosticautonomousrepositorylevel,
      title={MatchFixAgent: Language-Agnostic Autonomous Repository-Level Code Translation Validation and Repair}, 
      author={Ali Reza Ibrahimzada and Brandon Paulsen and Reyhaneh Jabbarvand and Joey Dodds and Daniel Kroening},
      year={2025},
      eprint={2509.16187},
      archivePrefix={arXiv},
      primaryClass={cs.SE},
      url={https://arxiv.org/abs/2509.16187}, 
}

\end{document}